\documentclass[aps,pra,twocolumn,superscriptaddress,amsmath,amssymb]{revtex4-1}
\usepackage[colorlinks,pdfusetitle,urlcolor=blue,citecolor=blue,linkcolor=blue,bookmarksnumbered,plainpages=false]{hyperref}
\usepackage{graphicx}
\usepackage{tikz}
\usepackage{braket}
\usepackage{nicefrac}

\newcommand{\laserFreq}{\omega_{\text{L}}}
\newcommand{\photonNumber}{N_{\text{L}}}
\newcommand{\B}{\mathbf}

\newcommand{\one}{\text{I}}
\newcommand{\two}{\text{II}}
\newcommand{\three}{\text{III}}

\newcommand{\rightInc}{\mathcal{R}}
\newcommand{\leftInc}{\mathcal{L}}

\newcommand{\trav}{\text{tr}}
\newcommand{\evan}{\text{ev}}

\newcommand{\imag}{\text{i}}

\newcommand{\intensity}{I}
\newcommand{\intensityClassical}{I_{\text{cl}}}

\newcommand{\dummy}{\chi}
\newcommand{\functionName}{S}

\newcommand{\groundState}{0}
\newcommand{\excitedState}{1}

\newcommand{\etaReplacement}{\tau}
\newcommand{\tauReplacement}{\kappa}

\newcommand{\down}{}
\newcommand{\up}{*}

\begin{document}

\title{The Casimir-Polder potential in the presence of a Fock state}

\author{Theodor Haug}
\affiliation{Physikalisches Institut, Albert-Ludwigs-Universit\"at Freiburg, 79104 Freiburg, Germany}

\author{Robert Bennett}
\affiliation{Physikalisches Institut, Albert-Ludwigs-Universit\"at Freiburg, 79104 Freiburg, Germany}

\author{Stefan Yoshi Buhmann}
\affiliation{Physikalisches Institut, Albert-Ludwigs-Universit\"at Freiburg, 79104 Freiburg, Germany}
\affiliation{Freiburg Institute for Advanced Studies, Albert-Ludwigs-Universit\"at Freiburg, Albertstra{\ss}e 19, 79104 Freiburg, Germany}

\date{\today}

\begin{abstract} Atom--surface forces using excited states have a host of compelling applications, including repulsive and lateral forces. However, such states can be fragile and difficult to prepare. Here we report an explicit normal-mode based calculation of the Casimir--Polder potential between a ground-state atom and a non-dispersive surface in the presence of an external quantised field. The potential we derive shares some features with that of excited-state Casimir-Polder forces even though we consider a ground--state atom. Our work provides a physically transparent and intuitive picture of driven Casimir-Polder potentials, as well as expanding on previous investigations by providing analytic results that fully include retardation, as well as being applicable for any choice of mutual alignment of the atom's dipole moment, the external field, and the surface normal.  \end{abstract}

\maketitle
\section{Introduction}
As technology progresses, the distances between construction element becomes continuously smaller, even reaching the nanometer regime \cite{Kim1999a,Browne2006}. At these scales the forces of the quantum vacuum can become significant \cite{Buks2001}, the most famous example being the Casimir force between two macroscopic bodies \cite{Casimir1948}. A closely related effect is the Casimir--Polder (CP) force between an atom and a macroscopic body \cite{Casimir1948a}, which is well-studied in a wide range of situations, using a variety of theoretical approaches \cite{McLachlan1963,Agarwal1975b,Wylie1984,Wylie1985,Eberlein2011,Buhmann2012BothBooks}. Some of the most compelling possibilities are found in systems where the atom is in an excited state \cite{Wylie1984,Wylie1985}, where forces can be repulsive or even lateral \cite{Rodriguez-Fortuno2015,Barcellona2018a}. Such effects are of great interest in cold-atom physics in order to realise tuneable interactions, however excited atomic states are often very short-lived. Recently, there has been growing interest in interplay between the CP effect and externally applied fields \cite{Perreault2008, Bender2010,Chang2014,Fuchs2017c}. The forces found there have some of the same desirable features as those for excited-state CP forces (the possibility of repulsion, for example), but, crucially, the atom is prepared in its ground state which sidesteps some experimental complications. 

The current literature on external field-modified CP forces can essentially be collected into two groups. The first is that taken by Ref.~\cite{Perreault2008}, where a simple, approximate model is built up in order to build intuition for the size and character of the effect. The second group consists of those works using more elaborate and sometimes opaque formalisms associated with field quantisation in dispersive and absorbing media. While very powerful, such approaches sometimes run into problems when attempting to describe physics that should at first sight be very simple --- see, for example, the way in which a laser-driving field has to be defined in Ref.~\cite{Fuchs2017c}. The goal of this work is to provide a suitable middle-ground between these two extremes. To do this, we will use an idealised but rigorous quantisation of the electromagnetic vacuum field near a non-dispersive medium (the so-called `triplet modes' \cite{Carniglia1971}), then use this to perturbatively calculate a CP force.

\begin{figure}\centering 
\includegraphics[width=0.7\columnwidth]{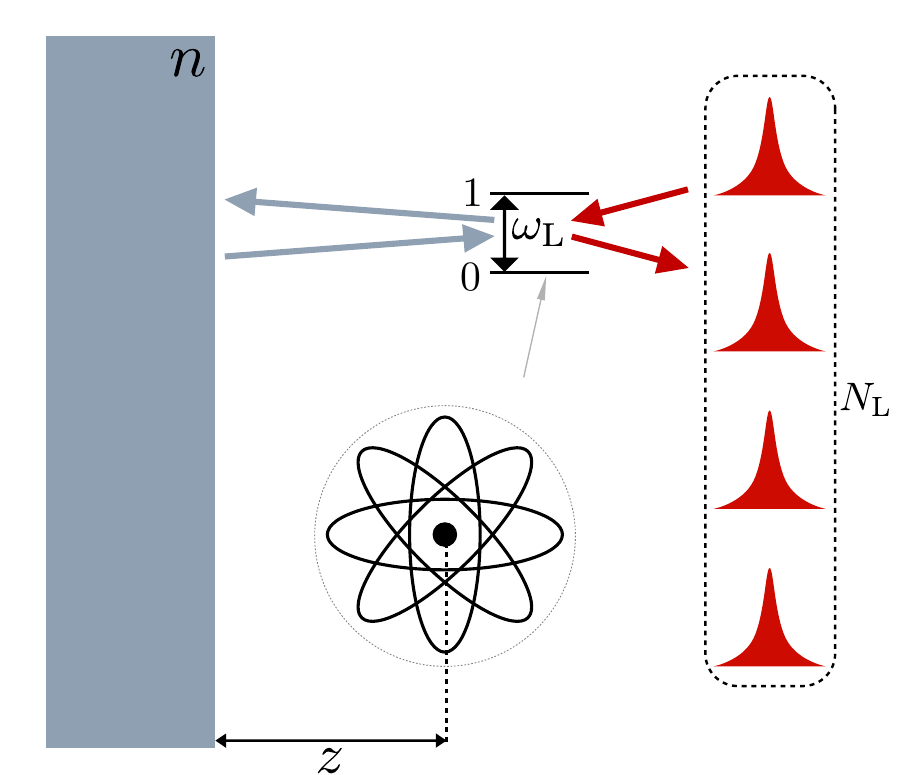}
\caption{ Schematic view of our setup. An atom is placed near an infinite planar surface of refractive index $n$, and allowed to interact with a light field in a number (Fock) state.}\label{AtomDiagram}
\end{figure}

 Aside from the methodological differences discussed above, we will consider a different type of external field to that used in previous work (e.g. \cite{Perreault2008, Chang2014, Fuchs2017c}). There the external field was classical, while here it is quantised and modelled as a Fock state with which the atom is allowed to exchange energy, as schematically indicated in Fig.~\ref{AtomDiagram}.
 In the limit where this external field contains a large number of excitations, we will show that our results reduce to those previously derived for a classical driving field.
 
 This paper is structured as follows: we begin in section \ref{SystemSection} by giving details of the system we intend to study, followed by section \ref{PerturbationSection} where we describe the perturbative calculation of an energy shift. Then, in section \ref{ResultsSection} we discuss our results and evaluate them for a selection of specific systems, followed by our conclusion in section \ref{ConclusionSection}.

\section{System}\label{SystemSection}

We consider a two-level atom, initially in its ground state $|\groundState \rangle$, subject to an external, quantized, monochromatic electric field $\hat{\B{E}}_\text{L}$., initially populated by $\photonNumber$ photons of frequency $\laserFreq$. As shown in Fig.~\ref{atom}, the atom is placed next to a medium with frequency-independent refractive index $n$, which occupies the half space $z<0$, while for $z>0$ we have vacuum. While of course this type of infinite planar surface does not exist in reality, it has proven itself to be a good approximation when the atom--surface distance is small relative to any geometric features, and in fact forms the basis of the proximity-force approximation (PFA) \cite{Derjaguin1934}, which is widely used in calculation of CP forces (see, for example, \cite{Lussange2012}).

\subsection{Field Quantization}\label{FQSection}

In order to derive vacuum-induced effects, we need to determine the vacuum electromagnetic field subject to the boundary conditions imposed by the surface. This will be achieved by the introduction of the triplet modes \cite{Carniglia1971}, where the electromagnetic field is separated into left and right-incident parts, which we will label $\leftInc$ and $\rightInc$, respectively. This field at spatial point $\B{r}$ and time $t$ can then be written in a very general form as;
\begin{equation}\label{ModesGeneralDefinition}
\hat{\textbf{E}}_\text{V}(\textbf{r},t)  =  \sum_{\lambda,\alpha}\int d^3 {k} \left[\textbf{f}^{\,\alpha}_{\textbf{k}\lambda}(\B{r},\omega)\hat{a}^\alpha_{\textbf{k}\lambda}e^{-\text{i}\omega t} +\mathrm{h.c.}\right]
\end{equation}
where $\textbf{f}^\alpha_{\textbf{k}\lambda}$ are position and frequency-dependent vector mode functions for each orthogonal polarization $\lambda =$ TE, TM and incidence direction $\alpha =\leftInc,\rightInc$. The set of bosonic operators $\hat{a}^\alpha_{\textbf{k}\lambda}$ and $\hat{a}^{\alpha\dagger}_{\textbf{k}\lambda}$ respectively create or annihilate an excitation of the quantised field with wave vector $\B{k}$, polarization $\lambda$ and incidence direction $\alpha$. The triplet modes are written by splitting each left or right-incident mode into incident (I), reflected (R) and transmitted (T) parts. 
\begin{align}
\textbf{f}^{\, \leftInc}_{\textbf{k}\lambda}(\textbf{r})&=\Theta(-z) \left[ \textbf{f}^{\, \leftInc,\text{I}}_{\textbf{k}\lambda}(\textbf{r})+\textbf{f}^{\, \leftInc,\text{R}}_{\textbf{k}\lambda}(\textbf{r})\right]+ \Theta(z)  \textbf{f}^{\, \leftInc,\text{T}}_{\textbf{k}\lambda}(\textbf{r})\label{LeftIncModes}\\
\textbf{f}^{\, \rightInc}_{\textbf{k}\lambda}(\textbf{r})&=\Theta(z) \left[ \textbf{f}^{\, \rightInc,\text{I}}_{\textbf{k}\lambda}(\textbf{r})+\textbf{f}^{\, \rightInc,\text{R}}_{\textbf{k}\lambda}(\textbf{r})\right]+ \Theta(-z)  \textbf{f}^{\, \rightInc,\text{T}}_{\textbf{k}\lambda}(\textbf{r})\label{RightIncModes}
\end{align}
where $\Theta(z)$ is the Heaviside step function and we have dropped frequency arguments for brevity. Here and throughout we use a system of units where $\hbar = c =\varepsilon_0 = 1$, unless otherwise specified.  In practice it is useful to package the incident and reflected parts of \eqref{LeftIncModes} and \eqref{RightIncModes} together by defining;
\begin{equation}
\textbf{f}^{\,\alpha,\text{IR}}_{\textbf{k}\lambda}(\textbf{r})=\textbf{f}^{\,\alpha,\text{I}}_{\textbf{k}\lambda}(\textbf{r})+\textbf{f}^{\,\alpha ,\text{R}}_{\textbf{k}\lambda}(\textbf{r})\,.
\end{equation}
The modes for the right hand side of the interface $(z>0)$ are given by \cite{Carniglia1971};
\begin{equation}\label{IncAndReflModes}
\textbf{f}^{\, \rightInc,\text{IR}}_{\textbf{k}\lambda}(\textbf{r},\omega)=-\text{i}\sqrt{\frac{\omega}{2(2\pi)^3}}\left[e^{\text{i}\textbf{k}\cdot\textbf{r}}\textbf{e}_{\textbf{k}\lambda}+R^\mathcal{R}_{\textbf{k}\lambda}e^{\text{i}\bar{\textbf{k}}\cdot\textbf{r}}\bar{\textbf{e}}_{\textbf{k}\lambda}\right] 
\end{equation}
and
\begin{equation}\label{TransModes}
\textbf{f}^{\,\leftInc,\text{T}}_{\textbf{k}\lambda}(\textbf{r},\omega) =-\text{i}\sqrt{\frac{\omega}{2(2\pi)^3}}\frac{1}{n}T^\mathcal{L}_{\textbf{k}\lambda}e^{\text{i}\textbf{k}\cdot\textbf{r}}\textbf{e}_{\textbf{k}\lambda}
\end{equation}
In these expressions the wave-vector $\textbf{k}$ has been decomposed into parts parallel and perpendicular to the surface, respectively denoted $\B{k}_{\|}$ and $k_z$.

\begin{figure}
\centering
\includegraphics[width=0.75\columnwidth]{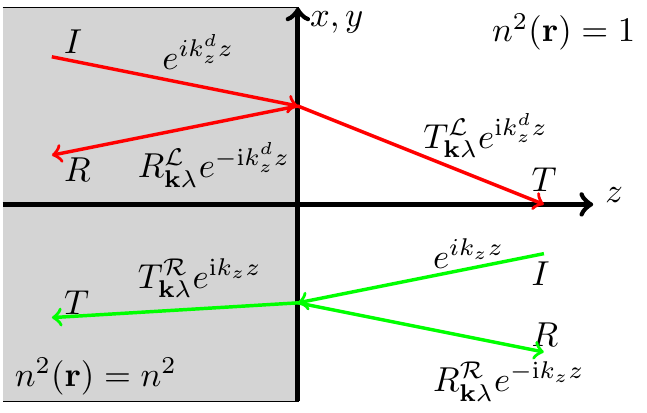}
\caption{Triplet mode functions}\label{atom}
\end{figure}

The vectors $\textbf{e}_{\textbf{k}\lambda}$ are unit polarization vectors indexed by wave-vector $\B{k}$ and polarization $\lambda$. Barred quantities here and throughout are obtained from unbarred ones via reflection $k_z \to -k_z$. The coefficients $R^\rightInc_{\textbf{k}\lambda}$ and $T^\leftInc_{\textbf{k}\lambda}$ are the reflection and transmission amplitudes relevant to each $\B{k}$ and $\lambda$, as illustrated in Fig.~\eqref{atom}. 
These turn out to be the well-known Fresnel coefficients as can be found by demanding that the mode functions either side of the interface should obey standard Maxwell boundary conditions \cite{Carniglia1971}. These are given explicitly by;
\begin{align}\label{reflection}
R^\rightInc_{\textbf{k}\text{TE}} &= \frac{k_z-\sqrt{n^2k_z^2+k_{\|}^2(n^2-1)}}{k_z+\sqrt{n^2k_z^2+k_{\|}^2(n^2-1)}}\equiv \frac{k_z-k_z^d}{k_z+k_z^d}\notag,
\\
R^\rightInc_{\textbf{k}\text{TM}} &= \frac{n^2k_z-\sqrt{n^2k_z^2+k_{\|}^2(n^2-1)}}{n^2k_z+\sqrt{n^2k_z^2+k_{\|}^2(n^2-1)}}\equiv \frac{n^2 k_z-k_z^d}{n^2 k_z+k_z^d}
\end{align}
where $k_z^d = \sqrt{n^2k_z^2+k_{\|}^2(n^2-1)}$ (with $k_\parallel=|\B{k}_\parallel|$) is the perpendicular component of the wave vector inside the medium. 
\subsection{External field}

Having set up the vacuum field, we now need to consider the external driving. We model this as a monochromatic quantised electric field, given by;
\begin{equation}
\hat{\textbf{E}}_\text{L}(\textbf{r},t) = \textbf{g}(\textbf{r},\laserFreq )e^{-\text{i}\laserFreq t}\hat{a}_\text{L} +\textbf{g}^*(\textbf{r},\laserFreq )e^{\text{i}\laserFreq t}\hat{a}^{\dagger}_\text{L},
\end{equation}
where $\hat{a}_\text{L}$ and  $\hat{a}^{\dagger}_\text{L}$ are the corresponding annihilation and creation operators. The only significant difference between this and a single mode of the vacuum field in free space is that the external field has an arbitrary amplitude given by $\textbf{E}_0=(E_x,E_y,E_z)^T$, from which we define the mode function for the external field;
\begin{equation}
\textbf{g}(\textbf{r},\omega)=-\text{i}\sqrt{\frac{\omega}{2(2\pi)}}\textbf{E}_0e^{\text{i}\textbf{k}_\text{L}\cdot\textbf{r}}.
\end{equation}
with $\laserFreq =  |\B{k}_\text{L}|$. 

\section{Perturbation theory}\label{PerturbationSection}

We model the interaction of the atom with the two fields via the electric dipole Hamiltonian $H_\text{AF}$;
\begin{equation}
\hat{H}_\text{AF} =-\hat{\textbf{d}}\cdot(\hat{\textbf{E}}_\text{V}+\hat{\textbf{E}}_\text{L}),
\end{equation}	
 where $\hat{\textbf{d}}$ is the electric dipole operator of the atom that couples its eigenstates. For the total Hamiltonian we then have;
 \begin{equation}\label{WholeHamiltonian}
H= a_\text{L}^\dagger a^{\vphantom{\dagger}}_{\text{L}}+ \sum_{\lambda,\alpha}\int d^3 {k}\,\hat{a}^{\alpha\dagger}_{\textbf{k}\lambda,\alpha}\hat{a}^{\alpha \vphantom{\dagger}}_{\textbf{k}\lambda} + \sum_{i=\groundState,\excitedState} E_i \ket{i}\bra{i}+\hat{H}_\text{AF} 
\end{equation}

We aim to calculate the energy shift caused by treating $\hat{H}_\text{AF} $ as a perturbation the Hamiltonian \eqref{WholeHamiltonian}. Since we are interested in effects which are surface \emph{and} external field dependent, we will keep only the terms which depend on both the vacuum field (which mediates the interaction with the surface) and the applied field. The resulting energy shift is then the CP potential, subject to our chosen external field.

 The interaction of the atom with the vacuum field proceeds via emission and reabsorption of a virtual photon, meaning that the lowest vacuum-dependent contributions appear in second-order perturbation theory. Similarly, the interaction with the external field proceeds either via emission and reabsorption of a photon, or absorption followed by reemission. In either case, such a process has its leading order contribution in second-order perturbation theory. Since we want to study the combined effects of both fields, we need to use fourth order perturbation theory. Thus our energy shift will be extracted from;
\begin{equation}\label{shift1}
\Delta E =\!\!\! \sum_{\substack{\one,\two,\three\neq \phi}}\!\!\! \!\!\frac{\bra{\phi} \!\hat{H}_\text{AF} \! \ket{\three}\!\bra{\three}\!\hat{H}_\text{AF}\! \ket{\two}\!\bra{\two}\!\hat{H}_\text{AF} \!\ket{\one}\!\bra{\one}\!\hat{H}_\text{AF} \ket{\phi}}{(E_\phi - E_\one)(E_\phi - E_\two)(E_\phi - E_\three)}
\end{equation}
where $\ket{\phi}$ is an initial state with energy $E_{\phi}$,  and $\ket{\one}$, $\ket{\two}$ and $ \ket{\three}$ are intermediate states to be found, with $E_\one$, $E_\two$ and $E_\three$ being their respective energies.

\begin{figure}[h!]\centering 
\includegraphics[width=\columnwidth]{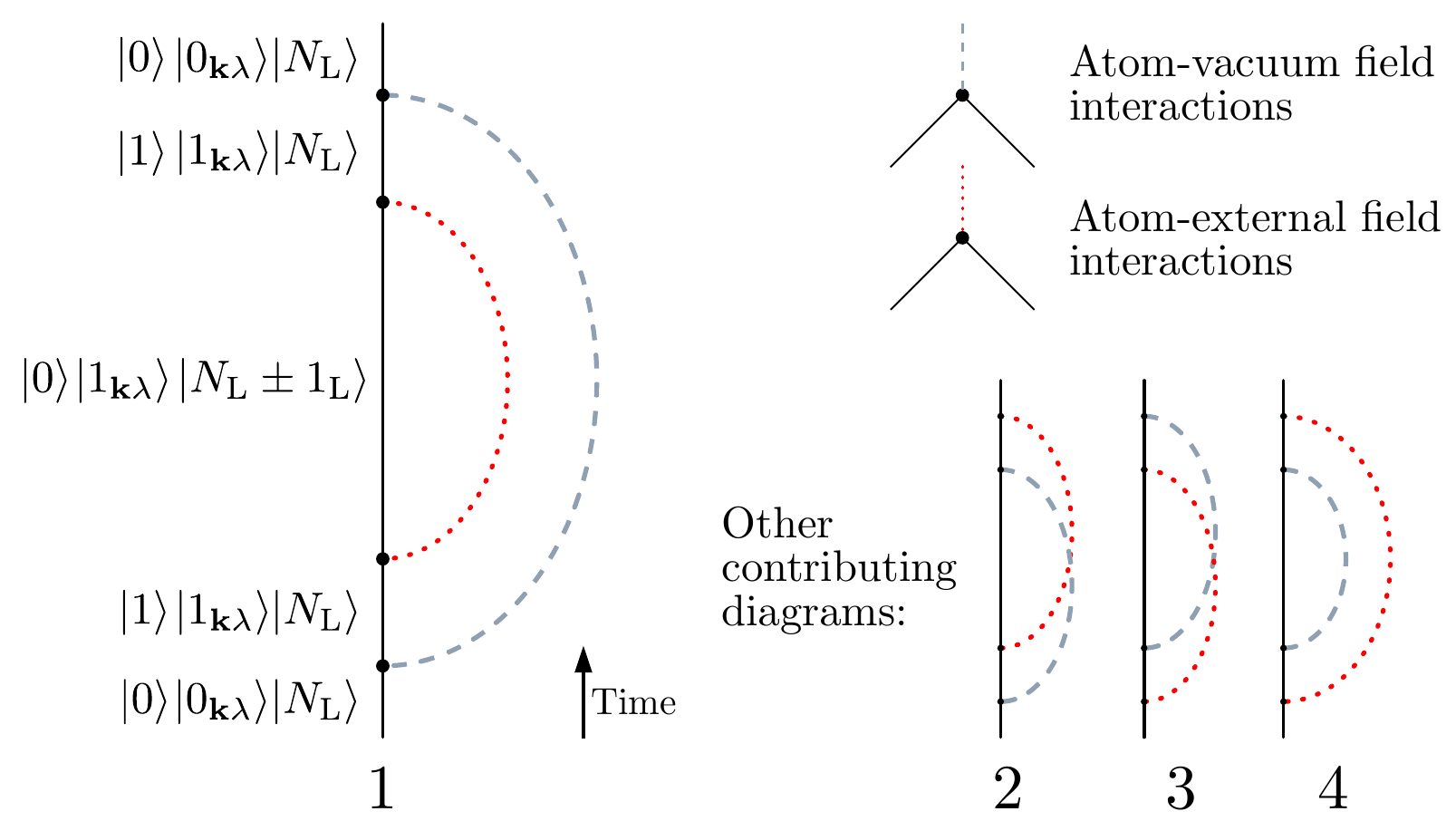}
\caption{Diagrammatic representation of the eight contributing terms in second-order perturbation theory. Each diagram above represents two contributions, as emission into or absorption from the external field are both possible directly from the initial state. The first diagram (1) is fully labelled, the others (2-4) can be labelled analogously.  }\label{diagrams}
\end{figure}

We choose the following initial state for our system:
\begin{equation}
\ket{\phi} = \ket{\groundState}\ket{0_{\mathbf{k}\lambda}} \ket{N_\text{L}},  \qquad E_{\phi} = E_\groundState + N_\text{L} \laserFreq
\end{equation}
corresponding to the atom being in state $\ket{\groundState}$, the empty vacuum field being represented as $\ket{0_{\mathbf{k}\lambda}}$, and the external bosonic field containing $N_\text{L}$ excitations. Proceeding through a lengthy evaluation of \eqref{shift1} using the techniques outlined in the next section, one finds eight relevant contributions, shown schematically in Fig.~\ref{diagrams}.
There it is seen that there are four different classes of diagram, each of which gives two contributions since the atom's first interaction with the external field could be either an emission into it ($\ket{N_\text{L}} \to \ket{N_\text{L}+1_\text{L}}$) or absorption from it ($\ket{N_\text{L}} \to \ket{N_\text{L}-1_\text{L}}$). In the next section we discuss in detail the evaluation of diagram 1 in Fig.~\ref{diagrams}, then simply quote the results for the others as these follow exactly analogously.

\subsection{Diagram 1}

We are now in a position to calculate the contribution of the first diagram to the energy shift. For this diagram, we have $\ket{\one}=\ket{\excitedState}\ket{1_{\textbf{k}\lambda}}\ket{\photonNumber}$, giving for the rightmost factor in \eqref{shift1};
\begin{align}
\bra{\one}\!\hat{H}_\text{AF} \ket{\phi}&=\bra{\photonNumber} \bra{1_{\textbf{k}\lambda}}\bra{\excitedState}\hat{\textbf{d}}\cdot(\hat{\textbf{E}}_\text{V}+\hat{\textbf{E}}_\text{L})\ket{\groundState}\ket{0_{\textbf{k}\lambda}}\ket{\photonNumber}\notag \\
&= \B{f}^*_{\textbf{k}\lambda}(\B{r},\omega) \cdot \B{d}^{\up} e^{\text{i}\omega t}\,.
\end{align}
where $\B{d}^{\up} \equiv \braket{1|\hat{\B{d}}|0}$. Proceeding analogously for the next factor to the left, we need the second intermediate state $\ket{\two}=\ket{\groundState}\ket{1_{\textbf{k}\lambda}}\ket{\photonNumber\pm 1_\text{L}}$ ;
\begin{equation}\label{trans2}
\bra{\two} \hat{H}_\text{AF} \ket{\one} =\sqrt{\photonNumber^{\pm}} \,\B{g} (\textbf{r},\laserFreq ) \cdot \B{d}^{\down}e^{-\text{i}\laserFreq t}.
\end{equation}
where we have defined 
\begin{equation}
\photonNumber^+ = \photonNumber + 1, \qquad \photonNumber^- = \photonNumber \, . 
\end{equation}
 Calculating the other factors in the numerator of \eqref{shift1} in the same way and simplifying the energy denominators, we find;
\begin{multline}\label{shift2}
\Delta E^{1\pm} = -\photonNumber^{\pm}\sum_{n,m,p,q}d^{\down}_nd^{\up}_md^{\down}_pd^{\up}_q\\\times\sum_{\lambda,\alpha}\mathcal{P} \int d^3 k\frac{f^\alpha_{n\textbf{k}\lambda}f^{\alpha*}_{q\textbf{k}\lambda} g_m g_p^*}{(\omega_{\down} + \omega)^2(\omega\pm \laserFreq )}\end{multline}
where the subscripts $n,m,p,q$ denote Cartesian directions, we have defined $\omega_{\down}\equiv  E_\excitedState-E_\groundState>0$ and $\mathcal{P}$ denotes the Cauchy principal value. Furthermore, we have used a shorthand notation where $f^\alpha_{n\textbf{k}\lambda}\equiv f^\alpha_{n\textbf{k}\lambda}(\B{r},\omega)$ and $g_n\equiv g_n(\textbf{r},\laserFreq )$. To further simplify Eq.~(\ref{shift2}) we assume that the atom is hydrogen-like, corresponding to the following relation for the dipole moment;
\begin{equation}\label{deltadipole}
d_n^{\up}d_m^{\down}=|d_n|^2\delta_{nm},
\end{equation}
giving;
\begin{multline}\label{shiftdiagram1-2}
\Delta E^{1\pm}= -\photonNumber^{\pm}\sum_{n,m}\sum_{\lambda,\alpha}\mathcal{P}\int d^3 k\frac{|f^\alpha_{n\textbf{k}\lambda}|^2|g_m|^2|d_n|^2|d_m|^2}{(\omega_{\down} + \omega)^2(\omega \pm \laserFreq )}
\end{multline}

\begin{figure}[h]
\centering
\includegraphics[width=\columnwidth]{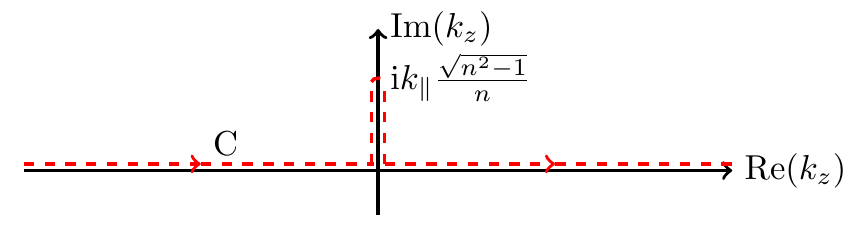}
\caption{The integration contour $C$ in the complex $k_z$ plane.}\label{contourC}
\end{figure}

To calculate quantities of the form $f^{\alpha\vphantom{*}}_{n\textbf{k}\lambda}f^{\alpha*}_{m\textbf{k}\lambda}$ one requires the outer product of the polarization vectors $\B{e}_{\B{k}\lambda}$ with themselves and their reflected counterparts $\bar{\B{e}}_{\B{k}\lambda}$. The resulting matrices are listed in Appendix \ref{PolVecApp}. In this particular case we need only the diagonal elements, using these one finds;
\begin{align}
&\sum_{n,\alpha} |f^\alpha_{n\textbf{k}\text{TE}}|^2|d_n|^2=  \int d^2\textbf{k}_{\|} \beta_{\textbf{k}\text{TE}}\left(k_y^2|d_x|^2+k_x^2|d_y|^2\right),\notag \\
&\sum_{n,\alpha} |f^\alpha_{n\textbf{k}\text{TM}}|^2|d_n|^2= \int d^2\textbf{k}_{\|} \beta_{\textbf{k}\text{TM}}\frac{1}{{k}^2{k}_{\|}^2} \notag\\
&\qquad\quad \times\left(-k_x^2k_z^2|d_x|^2-k_y^2k_z^2|d_y|^2+{k}_{\|}^4|d_z|^2\right),
\end{align}
where we have defined
\begin{equation}
\beta_{\textbf{k}\lambda}\equiv \int_{-\infty}^0 dk_z|e^{\text{i}\textbf{k}\cdot\textbf{r}} + R^\rightInc_{\textbf{k}\lambda}e^{\text{i}\bar{\textbf{k}}\cdot \B{r}}|^2 + \frac{1}{n^2}\int_0^{\infty}dk^d_z |T^\leftInc_{\textbf{k}\lambda}e^{\text{i}\textbf{k}\cdot\textbf{r}}|^2.
\end{equation}
This can be rewritten as a contour integral \cite{Eberlein2006,Bennett2012};
\begin{equation}\label{newbeta}
\beta_{\textbf{k}\lambda}= \int_Cdk_zR^\rightInc_{\textbf{k}\lambda}e^{2\text{i}k_zz}+\text{($z$-independent terms)},
\end{equation}
where the contour $C$ is shown in Fig.~\ref{contourC}.
\\
Since the energy shift is linear in $\beta_{\B{k}\lambda}$, we can isolate the surface-dependent part of the shift simply by discarding its $z$-independent terms. In approaches based on noise-current quantisation (see, for example, \cite{Fuchs2017c}), a similar isolation of surface-dependent effects is made by using the scattering Green's tensor rather than the whole Green's tensor.

Using the contour integral \eqref{newbeta} in Eq.~(\ref{shiftdiagram1-2}) we obtain for the surface-dependent part of the energy shift; 
\begin{align}\label{shiftdiagram1-3}
\notag\Delta & E^{1\pm}= -\frac{ \photonNumber^{\pm}\laserFreq }{4(2\pi)^4}(E_x^2|d_x|^2+E_y^2|d_y|^2+E_z^2|d_z|^2) \notag \\
&\times \int d^2\textbf{k}_{\|} \int_Cdk_z\frac{\omega e^{2\text{i}k_zz}}{(\omega_{\down} + \omega)^2(\omega \pm \laserFreq )}\notag  \\
&\times \bigg[\frac{R^\rightInc_{\textbf{k}\text{TE}}}{{k}_{\|}^2}\left(k_y^2|d_x|^2+k_x^2|d_y|^2\right)\notag \\& -\frac{R^\rightInc_{\textbf{k}\text{TM}}}{{\omega}^2{k}_{\|}^2}\left(k_x^2k_z^2|d_x|^2+k_y^2k_z^2|d_y|^2-{k}_{\|}^4|d_z|^2\right)\bigg]
\end{align}
Finally, we note from Eq.~\eqref{shiftdiagram1-3} that poles are encountered along the $k_z$ integration contour (recall that ${\omega^2 = k_z^2+k_\parallel^2}$), corresponding to resonant contributions where photons are absorbed from the external field. These are the terms we will concentrate on for the remainder of this work.

Recalling that $\omega_{\down}$ and $\laserFreq$ are positive by definition, it is apparent from \eqref{shiftdiagram1-3} that a pole is present at $\omega=\laserFreq$ for the case where there is a minus sign in the denominator. In terms of the integration variable $k_z$, the pole can be at one of three locations in the complex $k_z$ plane, depending on the relative values of $k_\parallel$ and $\laserFreq$. If $k_{\|}<\laserFreq $, the pole is on the real axis at;
\begin{equation}
k_z= \pm\sqrt{\laserFreq ^2-k_{\|}^2},
\end{equation}
corresponding to travelling wave solutions. By contrast, if $\laserFreq <k_{\|}$ the pole is found at.
\begin{equation}
k_z= \imag\sqrt{k_{\|}^2-\laserFreq ^2}.
\end{equation}
corresponding to evanescent modes, i.e. those which decay exponentially away from the interface. Here we have made explicit use of a physical constraint that implicitly enters into the definitions of the triplet modes \cite{Carniglia1971}, namely that $\text{Im}(k_z)>0$ in the region $z>0$. This ensures that the modes do not diverge as $z\to\infty$.

Since one pole is at purely imaginary $k_z$ and the other two poles are at purely real $k_z$, it is convenient to split the shift into two parts. Contributions originating from poles on the real axis will carry a subscript `$\trav$'(traveling-wave contribution), and similarly those originating from poles on the imaginary axis will carry a subscript `$\evan$' (evanescent-wave contribution).

\subsubsection{Poles on the real axis}

Using the Cauchy principal value prescription, one has two half-residue contributions. Using a shorthand $|d_{\|}|^2\equiv |d_x|^2+|d_y|^2$, we find for the total contribution of the real-axis poles;
\begin{multline}\label{shiftdiagram1-5}
\Delta E^{1}_\trav= -\text{Re}\frac{\text{i} \photonNumber\laserFreq ^4}{32\pi^2}\frac{1}{(\omega_{\down} + \laserFreq )^2}\\\times(E_x^2|d_x|^2+E_y^2|d_y|^2+E_z^2|d_z|^2) \int_0^{1} d\etaReplacement e^{2\text{i}\laserFreq \etaReplacement z}\\\times\left[|d^{\up}_{\|}|^2\left(R^\rightInc_{\laserFreq \text{TE}}-\etaReplacement^2R^\rightInc_{\laserFreq \text{TM}}\right)+2|d_z|^2(1-\etaReplacement^2)R^\rightInc_{\laserFreq \text{TM}}\right],
\end{multline}
where we transformed to polar coordinates via $k_x=k_\parallel \cos{\theta}$ and $k_y=k_{\|}\sin{\theta}$, carried out the trivial angular integral, then made another change of variable to $\etaReplacement = \sqrt{1-{k_{\|}^2}/{\laserFreq ^2}}$. The appearance of the real part is a consequence of one pole contribution being the complex conjugate of each other. 

\subsubsection{Pole on the imaginary axis}	
In a similar way, we find for the poles on the imaginary axis;

\begin{multline}\label{path1-evanfinal}
\Delta E^{1}_\evan= -\frac{ \photonNumber\laserFreq ^4}{32\pi^2}\frac{1}{(\omega_{\down} + \laserFreq )^2}\\\times(E_x^2|d_x|^2+E_y^2|d_y|^2+E_z^2|d_z|^2)\int_{0}^{\infty} d\tauReplacement e^{-2\laserFreq \tauReplacement z}\\\times\left[|d^{\up}_{\|}|^2\left(R^\rightInc_{\laserFreq\text{TE}}+\tauReplacement^2R^\rightInc_{\laserFreq\text{TM}}\right)+2|d_z|^2(1+\tauReplacement^2)R^\rightInc_{\laserFreq\text{TM}}\right],
\end{multline}
where we used a different integration variable, defined by $\tauReplacement = \sqrt{{k_{\|}^2}/{\laserFreq ^2}-1}$. Inspection of Eqs \eqref{shiftdiagram1-5}  and \eqref{path1-evanfinal}  reveals that they can be obtained from a single suitably defined function $\functionName_1$ (see Appendix \ref{IntegrandsApp}) via;
\begin{align}\label{ShortHandF}
\Delta E^{1}_\trav &= \text{Re}  \int_0^1 d\dummy \functionName_1(\dummy)\notag \\
\Delta E^{1}_\evan &=\int_0^\infty  \!\!d(\imag\dummy) \functionName_1(\imag\dummy)
\end{align}

The contributions from diagrams 2, 3 and 4 can also be written in the same way that, so that the total resonant energy shift can be written as;
\begin{align}
\label{ShortHandFTotTrav}\Delta E_\trav &= \sum_{i=1}^4\text{Re} \,   \int_0^1 d\dummy \functionName_i(\dummy)\equiv \text{Re} \,   \int_0^1 d\dummy \functionName_\text{tot}(\dummy) \\
\label{ShortHandFTotEv}\Delta E_\evan &=   \sum_{i=1}^4  \int_0^\infty d(\imag\dummy) \functionName_i(\imag\dummy)\equiv   \int_0^\infty d(\imag\dummy) \functionName_\text{tot}(\i\dummy)
\end{align}
We list the functions $\functionName_i$ corresponding to each individual diagram in Appendix \ref{IntegrandsApp}. Here we only report their total $\functionName_\text{tot}$;
\begin{widetext}
\begin{align}\label{FTot}
\functionName_\text{tot}(\dummy) = -\frac{ \photonNumber\laserFreq ^4}{32\pi^2 } \frac{e^{2\imag\laserFreq \dummy z}}{(\omega_{\down}^2 - \laserFreq ^2)^2} \Bigg\{{4\omega_{\down}^2}{}\Big[(E_x^2|d_x|^4+E_y^2|d_y|^4)\left(R^\rightInc_{\laserFreq \text{TE}} -\dummy^2R^\rightInc_{\laserFreq \text{TM}}\right)+2E_z^2|d_z|^4(1 -\dummy^2)R^\rightInc_{\laserFreq \text{TM}}\Big]& \notag \\ 
+2(\omega_{\down}^2+\laserFreq ^2)\bigg[\Big(E_z^2|d_z|^2|d^{\up}_\parallel|^2+(E_x^2+E_y^2)|d_x|^2|d_y|^2\Big)\left(R^\rightInc_{\laserFreq \text{TE}} -\dummy^2R^\rightInc_{\laserFreq \text{TM}}\right)& \notag \\
+2\Big(E_x^2|d_x|^2+E_y^2|d_y|^2\Big)|d_z|^2(1 -\dummy^2)R^\rightInc_{\laserFreq \text{TM}}\bigg]\Bigg\}&
\end{align}
\end{widetext}

\section{Results and discussion}\label{ResultsSection}

We now have the total resonant energy shift, given by;
\begin{equation}\label{GenResult}
\Delta E =\Delta E_\trav +\Delta E_\evan
\end{equation}
where $\Delta E_\trav$ and $\Delta E_\evan$ are defined by by Eqs.~\eqref{ShortHandFTotTrav}, \eqref{ShortHandFTotEv} and \eqref{FTot}. These lengthy expressions simplify significantly if some physical assumptions are made, for example if the external field and the dipole are both aligned in a direction parallel to the surface --- a case which we will distinguish by a superscript $\parallel$. Here one can choose $\textbf{E}_0=E_x \hat{\mathbf{x}}$ and $\B{d}={d}\hat{\mathbf{x}}$, finding;
\begin{multline}\label{shiftevan-2}
\Delta E^\parallel_\trav= -\frac{ \photonNumber\laserFreq ^4}{8\pi^2} \frac{\omega_{\down}^2}{(\omega_{\down}^2 - \laserFreq ^2)^2}\\\times\int_0^{\infty} d\tauReplacement E_x^2|d|^4e^{-2\laserFreq \tauReplacement z}\left(R^\rightInc_{\laserFreq \text{TE}}+\tauReplacement^2R^\rightInc_{\laserFreq \text{TM}}\right)
\end{multline}
and
\begin{multline}\label{shifttrav-2}
\Delta E^\parallel_\evan = -\text{Re}\frac{\text{i} \photonNumber\laserFreq ^4}{8\pi^2} \frac{\omega_{\down}^2}{(\omega_{\down}^2 - \laserFreq ^2)^2}\\\times\int_0^{1} d\etaReplacement E_x^2|d|^4e^{2\text{i}\laserFreq \etaReplacement z}\left(R^\rightInc_{\laserFreq \text{TE}}-\etaReplacement^2R^\rightInc_{\laserFreq \text{TM}}\right).
\end{multline}

 In the remainder of this section we will use the general results \eqref{shiftevan-2} and \eqref{shifttrav-2} to evaluate the energy shift for different materials.

\subsection{Perfect conductor, classical laser field}\label{perfectS}
As a consistency check of our general result \eqref{GenResult} we compare with \cite{Perreault2008}, where the CP force is calculated for an atom near a perfectly reflecting surface in the presence of a laser field. There, the external field and the dipole are parallel to each other and to the surface, so we can use the simplified expressions \eqref{shiftevan-2} and \eqref{shifttrav-2}. For a perfect reflector one has $R^\rightInc_{\textbf{k}\text{TE}} =-1$ and $R^\rightInc_{\textbf{k}\text{TM}}=1$, as can be verified from the reflection coefficients \eqref{reflection} by taking $n\to\infty$.

To model the laser field we use the correspondence principle to link the photon number $\photonNumber$ and the intensity of a classical field. The intensity of our field is;
\begin{equation}\label{intens}
\intensity = \frac12\langle \photonNumber|\hat{\B{E}}_\text{L}^2|\photonNumber\rangle = \frac{E_x^2\laserFreq }{8 \pi }(2\photonNumber+1) 
\end{equation} 
The classical limit should be recovered in the limit of large photon number $\photonNumber\gg 1$, under which conditions we have;
\begin{equation}\label{intens}
\intensity \approx \frac{E_x^2\photonNumber \laserFreq }{4 \pi} \equiv \intensityClassical
\end{equation} 
which we define as the intensity of the corresponding classical field. Inserting this into the general results given by Eq.~(\ref{shiftevan-2}) and (\ref{shifttrav-2}), we find the parallel-aligned, perfect conductor (PC) result $\Delta E^{\parallel\text{PC}} =\Delta E^{\parallel\text{PC}}_\trav +\Delta E^{\parallel\text{PC}}_\evan$ with;
\begin{equation}\label{shiftperfectevan}
\begin{aligned}
\Delta E^{\parallel\text{PC}}_\evan= -\frac{ \intensityClassical \alpha^2}{32\pi c\varepsilon_0^2}\frac{1}{z^3} \left(1-\frac{\zeta^2}{2}\right)
\end{aligned}
\end{equation}
and 
\begin{multline}\label{shiftperfecttrav}
\Delta E^{\parallel\text{PC}}_\trav=-\frac{\intensityClassical\alpha^2}{32\pi c \varepsilon_0^2}\frac{1}{z^3} \bigg[-1+\frac{\zeta^2}{2}\\
+(1-\zeta^2)\cos{\zeta}+\zeta \sin{\zeta }\bigg].
\end{multline}
where we have defined the dimensionless quantity $\zeta=2\laserFreq z/c$, made use of the polarizability $\alpha$;
\begin{equation}\label{pol}
\alpha \equiv \alpha (\laserFreq )= \frac{2}{\hbar}  \frac{\omega_{\down}|d|^2}{\omega_{\down}^2-\laserFreq ^2},
\end{equation}
and converted our expressions to S.I. units. In total we then have;
\begin{equation}\label{PCTotal}
\Delta E^{\parallel\text{PC}}=\frac{\alpha ^2\intensityClassical}{32 \pi  c \varepsilon_0^2 z^3} \left[\left(\zeta^2-1\right) \cos (\zeta)-\zeta \sin \zeta \right]
\end{equation}

We now use Eqs.~(\ref{shiftperfectevan}) and (\ref{shiftperfecttrav}) to explicitly evaluate the energy shift, for which we need to choose an atomic species and a laser field. We use $^{133}$Cs whose dominant transition is at a frequency of $\omega_{\down}= 1.55 \times 10^{14}$Hz with dipole moment of $d=5.85\times 10^{-29}$Cm \cite{NIST_ASD}. For the external field we choose a laser with a typical intensity $I=5{\text{W}}/{\text{cm}^2}$ and angular frequency $\laserFreq = 1.50 \times 10^{14}$Hz, which is in the mid-infrared. 
\begin{figure}[h]
\centering 
\includegraphics[width = \columnwidth]{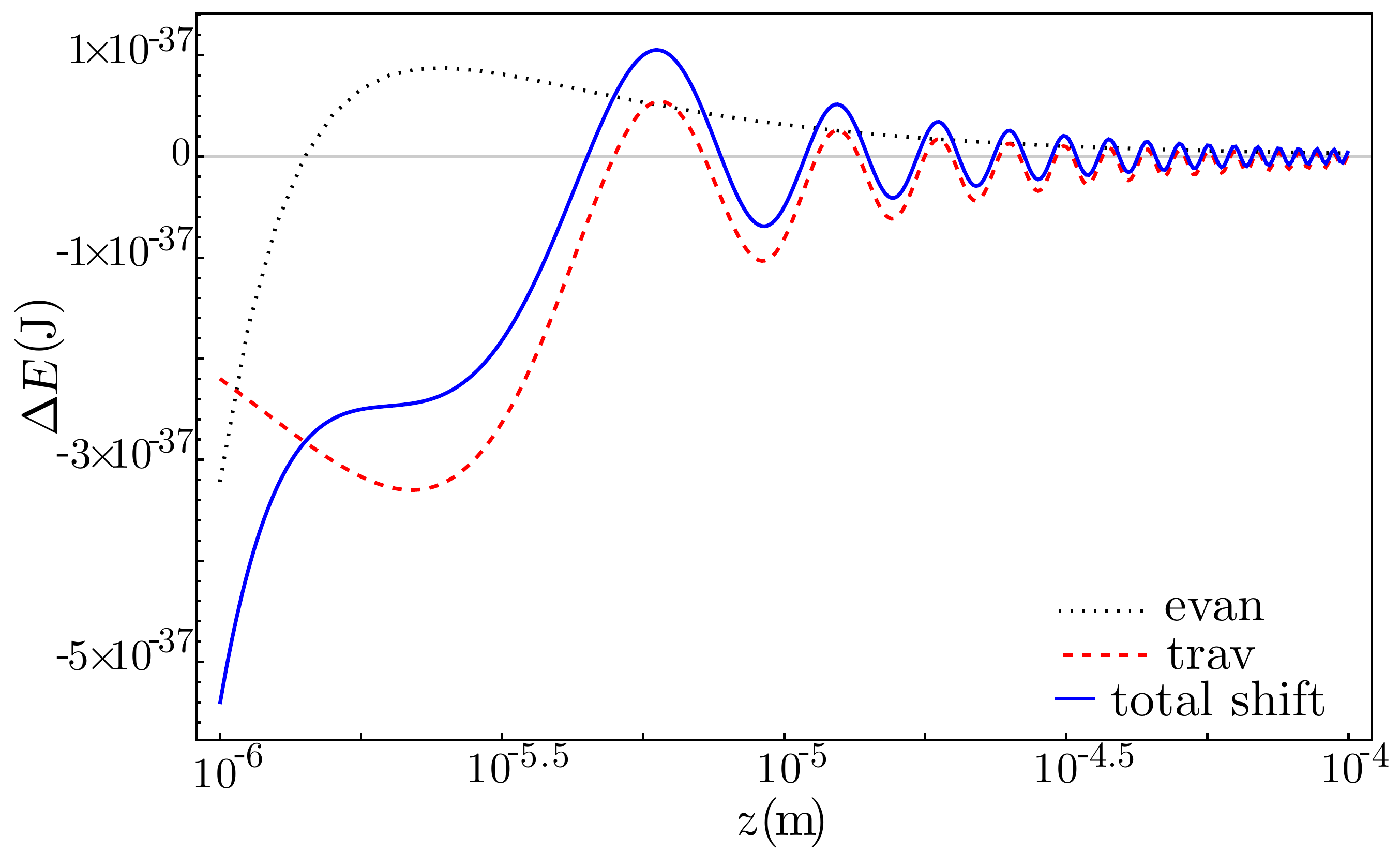}
\caption{Laser-driven CP potential of a $^{133}$Cs atom at the distance $z$ from a perfect conductor. Black dotted: evanescent-wave contribution; red-dashed: traveling-wave contribution; blue: total energy shift. }\label{Perfect}
\end{figure}\\
The results are shown in Fig.~\ref{Perfect} where we can see that the traveling-wave part has an oscillatory behaviour, meaning that the force can be attractive or repulsive depending on the exact position chosen. This is a remarkable feature of driven atom--surface forces --- usually one requires an atom in an initially excited state in order to realise such a potential, but here one can use a ground state atom which only visits its upper level in intermediate processes. 

The dimensionless quantity $\zeta/(2\pi)$ carries a physical meaning, which is most easily seen by writing;
\begin{equation}
\frac{\zeta}{2\pi} = \frac{1}{2\pi} \cdot  \frac{2\laserFreq z}{c} =\frac{2z/c}{2\pi/\laserFreq} 
\end{equation}
which is the ratio of the round-trip time for a photon of frequency $\laserFreq$ to travel from the atom to the surface and back again, and the time period of the external field. If we are in the non-retarded regime (${\zeta} \ll 2\pi $), the state of the atom will not have appreciably changed during photon exchange with the surface, so no oscillatory (phased-based) effects should appear. Indeed, expanding Eq.~\eqref{PCTotal} for small $\zeta$ one finds;
\begin{equation}
\begin{aligned}
\Delta E^{\parallel\text{PC}}({\zeta} \ll 2\pi)= -\frac{ \intensityClassical \alpha^2}{32\pi c\varepsilon_0^2}\frac{1}{z^3}.
\end{aligned}
\end{equation}
which agrees with Ref.~\cite{ Perreault2008}'s Eq.~(28). In the retarded (long-distance) limit we find; 
\begin{equation}
\begin{aligned}\label{RetResult}
\Delta E^{\parallel\text{PC}}({\zeta}\gg 2\pi)=\frac{\intensityClassical\alpha^2\laserFreq ^2}{8\pi c^3\varepsilon_0^2}\frac{\cos{({2\laserFreq z }/{c}})}{z}.
\end{aligned}
\end{equation}
In Ref.~\cite{ Perreault2008} it is claimed that retardation is included, but our result \eqref{RetResult} differs from theirs. To see why it is instructive to rewrite the shift \eqref{PCTotal} as;
\begin{equation}\label{TermsComparison}
\Delta E^{\parallel\text{PC}}=\frac{\intensityClassical \alpha ^2\laserFreq^3 }{8 \pi  c^4 \varepsilon_0^2 } \left[ -\frac{\cos \zeta}{\zeta^3}-\frac{\sin \zeta}{\zeta^2}+\frac{\cos \zeta}{\zeta} \right]\end{equation}
The first term is the result of \cite{Perreault2008}, where it is claimed that such a result includes retardation, encoded via the factor $\cos\zeta$. In the near-field limit, our results agree with \cite{Perreault2008} since the cosine factor becomes unity and the other terms are negligible. However, going to next-to-leading order in the non-retarded approximation (i.e. including a small retardation-dependent effect), one finds terms of order $1/\zeta$ originating from all three terms of Eq.~\eqref{TermsComparison}. In  \cite{Perreault2008} all terms of order $\zeta^{-1}$ and $\zeta^{-2}$ are discarded, which is valid but is inconsistent with including a cosine factor in the first term, since only the leading order contribution to that term (which is $\propto \cos(0)/\zeta^{-3}=1/\zeta^{-3}$) is insensitive to the discarding of terms of order $\zeta^{-1}$ and $\zeta^{-2}$. This means that the work of \cite{Perreault2008} in fact contains only results for the non-retarded regime where oscillatory results cannot be observed at leading order, so it is not surprising that our result \eqref{RetResult} differs from theirs.

\subsection{Imperfect reflection}
\begin{figure}\centering 
\includegraphics[width=0.5\textwidth]{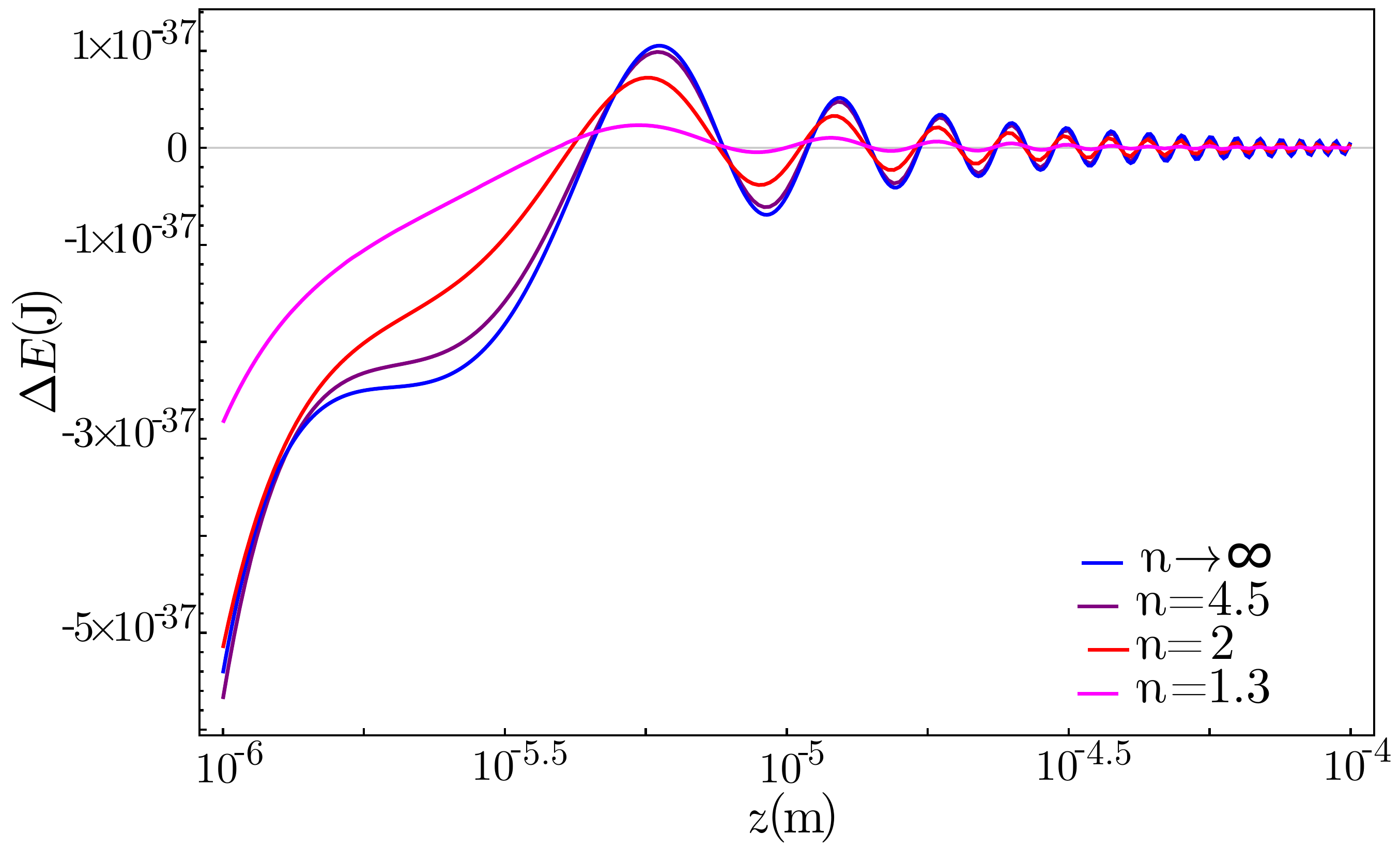}
\caption{Laser-driven CP potential of a $^{133}$Cs atom at distance $z$ from different materials. Blue: $n\rightarrow\infty$ (perfect conductor), purple: $n\approx4.5$ (e.g. sapphire), pink: $n=2$ (e.g.~sulphur in the visible spectrum), magenta: $n=1.3 $ (e.g.~water in the visible spectrum)}\label{differentN}
\end{figure}

The modes outlined in section \eqref{FQSection} apply for a dielectric half-space of refractive index $n$, with the the previous section dealing with the analytically solvable special case of a perfect reflector. For imperfect reflection (finite $n$), the integrals Eqs.~(\ref{shiftevan-2}) and (\ref{shifttrav-2}) become
\begin{multline}\label{shiftevanmat}
\Delta E_\evan^\parallel = -\frac{\intensityClassical\alpha^2\laserFreq ^3}{8\pi}\int_0^{\infty} d\tauReplacement e^{-2\laserFreq\tauReplacement z}\\\times\left(\frac{\tauReplacement - \sqrt{1 - n^2 + \tauReplacement^2}}{\tauReplacement + \sqrt{1 - n^2 + \tauReplacement^2}}+\tauReplacement^2\frac{n^2\tauReplacement - \sqrt{1 - n^2 + \tauReplacement^2}}{n^2\tauReplacement + \sqrt{1 - n^2 + \tauReplacement^2}}\right)
\end{multline}
and
\begin{multline}\label{shifttravmat}
\Delta E_\trav^\parallel  = -\text{Re}\frac{\text{i}\intensityClassical \alpha^2\laserFreq ^3}{8\pi }\int_0^{1} d\etaReplacement e^{2\text{i}\laserFreq \etaReplacement z}\\\times\left(\frac{\etaReplacement - \sqrt{1 - n^2 + \etaReplacement^2}}{\etaReplacement + \sqrt{1 - n^2 + \etaReplacement^2}}-\etaReplacement^2\frac{n^2\etaReplacement - \sqrt{1 - n^2 + \etaReplacement^2}}{n^2\etaReplacement + \sqrt{1 - n^2 + \etaReplacement^2}}\right)
\end{multline}
which must be solved numerically. The resulting energy shifts for a range of $n$ are shown in Fig.~\ref{differentN}.
It is observed there that finite $n$ causes relatively minor corrections to the perfect reflector result, except for the case $n=1.3$ which could be considered as a dilute medium. 

\section{Conclusion}\label{ConclusionSection}
In this paper we have used explicit and physically transparent perturbation theory to calculate a general result for the surface-- and external field--dependent energy shift of a two-level atom interacting with a populated photon state in near a dielectric half-space. These equations allow one to calculate the resonant energy shift of a two-level atom with arbitrary mutual alignment of the dipole, applied field, and surface normal. Our results were calculated using the example of a non-dispersive medium, but since we concentrated on resonant terms (those which depend on the medium's response at a single frequency), our results also can also be applied to dispersive media simply by taking the permittivity at the relevant frequency.  We have confirmed our general results by showing that they reduce to and expand upon the perfect reflector results shown previously by others, and shown some illustrative examples for imperfect reflection. Our results show that the qualitative behaviour of excited--state Casimir--Polder forces can be reproduced via external driving, which has applications in trapping and control of cold atoms. The intricate dependence of the force on the atom--surface distance could be used in future as a sensitive probe of surface structure and response. 

\begin{acknowledgments}
The authors acknowledge financial support from the Deutsche Forschungsgemeinschaft (grant BU 1803/3-1476). R.B. additionally acknowledges from the Alexander von Humboldt Foundation, and S.Y.B acknowledges support from the Freiburg Institute for Advanced Studies (FRIAS).
\end{acknowledgments}
\appendix

\section{Polarization matrices}\label{PolVecApp}

A suitable choice of polarization vectors is;
\begin{equation}
\textbf{e}_{\textbf{k}\text{TE}}=\frac{1}{{k}_{\|}}\begin{pmatrix} k_y \\ -k_x \\ 0 \end{pmatrix},  \qquad \textbf{e}_{{k}\text{TM}}=\frac{1}{{k}{k}_{\|}}\begin{pmatrix} k_xk_z \\ k_yk_z \\ -{k}_{\|}^2 \end{pmatrix} .
\end{equation}
In the main text we need the dyadic product of these with their reflected counterparts $\bar{\textbf{e}}_{\textbf{k}\lambda}$. Such products are given explicitly by;
\begin{align*}
\notag\textbf{e}_{\textbf{k}\text{TE}} \otimes\bar{\textbf{e}}_{\textbf{k}\text{TE}} =& \frac{1}{{k}_{\|}^2}
\begin{pmatrix}
k_y^2 & -k_xk_y & 0 \\
-k_xk_y & k_x^2 & 0 \\
0 &0&0\\
\end{pmatrix}
\\
\textbf{e}_{\textbf{k}\text{TM}} \otimes \bar{\textbf{e}}_{\textbf{k}\text{TM}} =& \frac{1}{{k}_{\|}^2{k}^2}
\begin{pmatrix}
-k_x^2k_z^2 & -k_xk_yk_z^2 &- k_xk_z{k}_{\|}^2 \\
 -k_xk_yk_z^2 &-k_y^2k_z^2 &  -k_yk_z{k}_{\|}^2 \\
 k_xk_z{k}_{\|}^2 & k_yk_z{k}_{\|}^2&{k}_{\|}^4\\
\end{pmatrix}.
\end{align*}

\section{Integrands}\label{IntegrandsApp}

The integrands of \eqref{ShortHandFTotTrav} and \eqref{ShortHandFTotEv} required for calculating the contribution of diagram 1 (see Fig.~\ref{diagrams}) to the shift \eqref{shift1} are found from;
\begin{align}
\functionName_1(\dummy) &= -\frac{\text{i} \photonNumber\laserFreq ^4}{32\pi^2 }\frac{e^{2\text{i}\laserFreq \dummy z}}{(\omega_{\down} + \laserFreq )^2}\notag \\ 
&\quad \times (E_x^2|d_x|^2+E_y^2|d_y|^2+E_z^2|d_z|^2) \notag \\ 
&\quad\quad \times\Big[|d^{}_{\|}|^2\left(R^\rightInc_{\laserFreq \text{TE}}-\dummy^2R^\rightInc_{\laserFreq \text{TM}}\right)\notag \\
&\quad\quad\quad \quad\qquad\quad+2|d_z|^2(1-\dummy^2)R^\rightInc_{\laserFreq \text{TM}}\Big].
\end{align}
It turns out that diagrams 2 and 3 make equal contributions to \eqref{shift1}, as they must do since they are exactly the same process in reverse time order. We find; \begin{align}
&\functionName_2(\dummy)  = -\frac{\text{i} \photonNumber\laserFreq ^4}{32\pi^2}\frac{e^{2\text{i}\laserFreq \dummy z}}{(\omega_{\down} + \laserFreq )(\omega_{\down} - \laserFreq )} \notag \\ 
& \times \Big[ (E_x^2|d_x|^4+E_y^2|d_y|^4)\left(R^\rightInc_{\laserFreq \text{TE}}-\dummy^2R^\rightInc_{\laserFreq \text{TM}}\right)\notag \\
&\qquad \qquad+ 2E_z^2|d_z|^4(1-\dummy^2)R^\rightInc_{\laserFreq \text{TM}} \Big]=\functionName_3(\dummy)
\end{align}
Finally the contribution of diagram four is found from;
\begin{align}
\functionName_4(\dummy) &= -\frac{\text{i} \photonNumber\laserFreq ^4}{32 \pi^2} \frac{e^{2\text{i}\laserFreq \dummy z}}{(\omega_{\down} - \laserFreq )^2}\notag \\ 
&\quad\times(E_x^2|d_x|^2+E_y^2|d_y|^2+E_z^2|d_z|^2)\notag \\
&\quad\quad\times\Big[|d^{}_{\|}|^2\left(R^\rightInc_{\laserFreq \text{TE}}-\dummy^2R^\rightInc_{\laserFreq \text{TM}}\right)\notag \\
&\qquad \qquad \qquad\quad +2|d_z|^2(1-\dummy^2)R^\rightInc_{\laserFreq \text{TM}}\Big].
\end{align}


\begin{thebibliography}{23}%
\makeatletter
\providecommand \@ifxundefined [1]{%
 \@ifx{#1\undefined}
}%
\providecommand \@ifnum [1]{%
 \ifnum #1\expandafter \@firstoftwo
 \else \expandafter \@secondoftwo
 \fi
}%
\providecommand \@ifx [1]{%
 \ifx #1\expandafter \@firstoftwo
 \else \expandafter \@secondoftwo
 \fi
}%
\providecommand \natexlab [1]{#1}%
\providecommand \enquote  [1]{``#1''}%
\providecommand \bibnamefont  [1]{#1}%
\providecommand \bibfnamefont [1]{#1}%
\providecommand \citenamefont [1]{#1}%
\providecommand \href@noop [0]{\@secondoftwo}%
\providecommand \href [0]{\begingroup \@sanitize@url \@href}%
\providecommand \@href[1]{\@@startlink{#1}\@@href}%
\providecommand \@@href[1]{\endgroup#1\@@endlink}%
\providecommand \@sanitize@url [0]{\catcode `\\12\catcode `\$12\catcode
  `\&12\catcode `\#12\catcode `\^12\catcode `\_12\catcode `\%12\relax}%
\providecommand \@@startlink[1]{}%
\providecommand \@@endlink[0]{}%
\providecommand \url  [0]{\begingroup\@sanitize@url \@url }%
\providecommand \@url [1]{\endgroup\@href {#1}{\urlprefix }}%
\providecommand \urlprefix  [0]{URL }%
\providecommand \Eprint [0]{\href }%
\providecommand \doibase [0]{http://dx.doi.org/}%
\providecommand \selectlanguage [0]{\@gobble}%
\providecommand \bibinfo  [0]{\@secondoftwo}%
\providecommand \bibfield  [0]{\@secondoftwo}%
\providecommand \translation [1]{[#1]}%
\providecommand \BibitemOpen [0]{}%
\providecommand \bibitemStop [0]{}%
\providecommand \bibitemNoStop [0]{.\EOS\space}%
\providecommand \EOS [0]{\spacefactor3000\relax}%
\providecommand \BibitemShut  [1]{\csname bibitem#1\endcsname}%
\let\auto@bib@innerbib\@empty
\bibitem [{\citenamefont {Kim}\ and\ \citenamefont {Lieber}(1999)}]{Kim1999a}%
  \BibitemOpen
  \bibfield  {author} {\bibinfo {author} {\bibfnamefont {P.}~\bibnamefont
  {Kim}}\ and\ \bibinfo {author} {\bibfnamefont {C.~M.}\ \bibnamefont
  {Lieber}},\ }\href {\doibase 10.1126/science.286.5447.2148} {\bibfield
  {journal} {\bibinfo  {journal} {Science}\ }\textbf {\bibinfo {volume}
  {286}},\ \bibinfo {pages} {2148} (\bibinfo {year} {1999})}\BibitemShut
  {NoStop}%
\bibitem [{\citenamefont {Browne}\ and\ \citenamefont
  {Feringa}(2006)}]{Browne2006}%
  \BibitemOpen
  \bibfield  {author} {\bibinfo {author} {\bibfnamefont {W.~R.}\ \bibnamefont
  {Browne}}\ and\ \bibinfo {author} {\bibfnamefont {B.~L.}\ \bibnamefont
  {Feringa}},\ }\href {\doibase 10.1038/nnano.2006.45} {\bibfield  {journal}
  {\bibinfo  {journal} {Nature Nanotechnology}\ }\textbf {\bibinfo {volume}
  {1}},\ \bibinfo {pages} {25} (\bibinfo {year} {2006})}\BibitemShut {NoStop}%
\bibitem [{\citenamefont {Buks}\ and\ \citenamefont {Roukes}(2001)}]{Buks2001}%
  \BibitemOpen
  \bibfield  {author} {\bibinfo {author} {\bibfnamefont {E.}~\bibnamefont
  {Buks}}\ and\ \bibinfo {author} {\bibfnamefont {M.~L.}\ \bibnamefont
  {Roukes}},\ }\href {\doibase 10.1103/PhysRevB.63.033402} {\bibfield
  {journal} {\bibinfo  {journal} {Physical Review B}\ }\textbf {\bibinfo
  {volume} {63}},\ \bibinfo {pages} {033402} (\bibinfo {year}
  {2001})}\BibitemShut {NoStop}%
\bibitem [{\citenamefont {Casimir}(1948)}]{Casimir1948}%
  \BibitemOpen
  \bibfield  {author} {\bibinfo {author} {\bibfnamefont {H.}~\bibnamefont
  {Casimir}},\ }\href {\doibase citeulike-article-id:8810715} {\bibfield
  {journal} {\bibinfo  {journal} {Proc. K. Ned. Akad.}\ }\textbf {\bibinfo
  {volume} {360}},\ \bibinfo {pages} {793} (\bibinfo {year}
  {1948})}\BibitemShut {NoStop}%
\bibitem [{\citenamefont {Casimir}\ and\ \citenamefont
  {Polder}(1948)}]{Casimir1948a}%
  \BibitemOpen
  \bibfield  {author} {\bibinfo {author} {\bibfnamefont {H.~B.~G.}\
  \bibnamefont {Casimir}}\ and\ \bibinfo {author} {\bibfnamefont
  {D.}~\bibnamefont {Polder}},\ }\href {\doibase 10.1103/PhysRev.73.360}
  {\bibfield  {journal} {\bibinfo  {journal} {Physical Review}\ }\textbf
  {\bibinfo {volume} {73}},\ \bibinfo {pages} {360} (\bibinfo {year}
  {1948})}\BibitemShut {NoStop}%
\bibitem [{\citenamefont {McLachlan}(1963)}]{McLachlan1963}%
  \BibitemOpen
  \bibfield  {author} {\bibinfo {author} {\bibfnamefont {A.~D.}\ \bibnamefont
  {McLachlan}},\ }\href {\doibase 10.1098/rspa.1963.0025} {\bibfield  {journal}
  {\bibinfo  {journal} {Proceedings of the Royal Society A: Mathematical,
  Physical and Engineering Sciences}\ }\textbf {\bibinfo {volume} {271}},\
  \bibinfo {pages} {387} (\bibinfo {year} {1963})}\BibitemShut {NoStop}%
\bibitem [{\citenamefont {Agarwal}(1975)}]{Agarwal1975b}%
  \BibitemOpen
  \bibfield  {author} {\bibinfo {author} {\bibfnamefont {G.~S.}\ \bibnamefont
  {Agarwal}},\ }\href {\doibase 10.1103/PhysRevA.11.243} {\bibfield  {journal}
  {\bibinfo  {journal} {Physical Review A}\ }\textbf {\bibinfo {volume} {11}},\
  \bibinfo {pages} {243} (\bibinfo {year} {1975})}\BibitemShut {NoStop}%
\bibitem [{\citenamefont {Wylie}\ and\ \citenamefont {Sipe}(1984)}]{Wylie1984}%
  \BibitemOpen
  \bibfield  {author} {\bibinfo {author} {\bibfnamefont {J.~M.}\ \bibnamefont
  {Wylie}}\ and\ \bibinfo {author} {\bibfnamefont {J.~E.}\ \bibnamefont
  {Sipe}},\ }\href {\doibase 10.1103/PhysRevA.30.1185} {\bibfield  {journal}
  {\bibinfo  {journal} {Physical Review A}\ }\textbf {\bibinfo {volume} {30}},\
  \bibinfo {pages} {1185} (\bibinfo {year} {1984})}\BibitemShut {NoStop}%
\bibitem [{\citenamefont {Wylie}\ and\ \citenamefont {Sipe}(1985)}]{Wylie1985}%
  \BibitemOpen
  \bibfield  {author} {\bibinfo {author} {\bibfnamefont {J.~M.}\ \bibnamefont
  {Wylie}}\ and\ \bibinfo {author} {\bibfnamefont {J.~E.}\ \bibnamefont
  {Sipe}},\ }\href {\doibase 10.1103/PhysRevA.32.2030} {\bibfield  {journal}
  {\bibinfo  {journal} {Physical Review A}\ }\textbf {\bibinfo {volume} {32}},\
  \bibinfo {pages} {2030} (\bibinfo {year} {1985})}\BibitemShut {NoStop}%
\bibitem [{\citenamefont {Eberlein}\ and\ \citenamefont
  {Zietal}(2011)}]{Eberlein2011}%
  \BibitemOpen
  \bibfield  {author} {\bibinfo {author} {\bibfnamefont {C.}~\bibnamefont
  {Eberlein}}\ and\ \bibinfo {author} {\bibfnamefont {R.}~\bibnamefont
  {Zietal}},\ }\href {\doibase 10.1103/PhysRevA.83.052514} {\bibfield
  {journal} {\bibinfo  {journal} {Physical Review A}\ }\textbf {\bibinfo
  {volume} {83}},\ \bibinfo {pages} {052514} (\bibinfo {year}
  {2011})}\BibitemShut {NoStop}%
\bibitem [{\citenamefont {Buhmann}(2012)}]{Buhmann2012BothBooks}%
  \BibitemOpen
  \bibfield  {author} {\bibinfo {author} {\bibfnamefont {S.~Y.}\ \bibnamefont
  {Buhmann}},\ }\href@noop {} {\emph {\bibinfo {title} {{Dispersion
  Forces}}}},\ \bibinfo {series} {Springer Tracts in Modern Physics}, Vol.\
  \bibinfo {volume} {247}\ (\bibinfo  {publisher} {Springer},\ \bibinfo
  {address} {Berlin, Heidelberg},\ \bibinfo {year} {2012})\ p.\ \bibinfo
  {pages} {330}\BibitemShut {NoStop}%
\bibitem [{\citenamefont {Rodr{\'{i}}guez-Fortu{\~{n}}o}\ \emph
  {et~al.}(2015)\citenamefont {Rodr{\'{i}}guez-Fortu{\~{n}}o}, \citenamefont
  {Engheta}, \citenamefont {Mart{\'{i}}nez},\ and\ \citenamefont
  {Zayats}}]{Rodriguez-Fortuno2015}%
  \BibitemOpen
  \bibfield  {author} {\bibinfo {author} {\bibfnamefont {F.~J.}\ \bibnamefont
  {Rodr{\'{i}}guez-Fortu{\~{n}}o}}, \bibinfo {author} {\bibfnamefont
  {N.}~\bibnamefont {Engheta}}, \bibinfo {author} {\bibfnamefont
  {A.}~\bibnamefont {Mart{\'{i}}nez}}, \ and\ \bibinfo {author} {\bibfnamefont
  {A.~V.}\ \bibnamefont {Zayats}},\ }\href {\doibase 10.1038/ncomms9799}
  {\bibfield  {journal} {\bibinfo  {journal} {Nature Communications}\ }\textbf
  {\bibinfo {volume} {6}},\ \bibinfo {pages} {8799} (\bibinfo {year}
  {2015})}\BibitemShut {NoStop}%
\bibitem [{\citenamefont {Barcellona}\ \emph {et~al.}(2018)\citenamefont
  {Barcellona}, \citenamefont {Bennett},\ and\ \citenamefont
  {Buhmann}}]{Barcellona2018a}%
  \BibitemOpen
  \bibfield  {author} {\bibinfo {author} {\bibfnamefont {P.}~\bibnamefont
  {Barcellona}}, \bibinfo {author} {\bibfnamefont {R.}~\bibnamefont {Bennett}},
  \ and\ \bibinfo {author} {\bibfnamefont {S.~Y.}\ \bibnamefont {Buhmann}},\
  }\href {http://arxiv.org/abs/1802.09234} {\  (\bibinfo {year} {2018})},\
  \Eprint {http://arxiv.org/abs/1802.09234} {arXiv:1802.09234} \BibitemShut
  {NoStop}%
\bibitem [{\citenamefont {Perreault}\ \emph {et~al.}(2008)\citenamefont
  {Perreault}, \citenamefont {Bhattacharya}, \citenamefont {Lonij},\ and\
  \citenamefont {Cronin}}]{Perreault2008}%
  \BibitemOpen
  \bibfield  {author} {\bibinfo {author} {\bibfnamefont {J.~D.}\ \bibnamefont
  {Perreault}}, \bibinfo {author} {\bibfnamefont {M.}~\bibnamefont
  {Bhattacharya}}, \bibinfo {author} {\bibfnamefont {V.~P.~A.}\ \bibnamefont
  {Lonij}}, \ and\ \bibinfo {author} {\bibfnamefont {A.~D.}\ \bibnamefont
  {Cronin}},\ }\href {\doibase 10.1103/PhysRevA.77.043406} {\bibfield
  {journal} {\bibinfo  {journal} {Physical Review A}\ }\textbf {\bibinfo
  {volume} {77}},\ \bibinfo {pages} {043406} (\bibinfo {year}
  {2008})}\BibitemShut {NoStop}%
\bibitem [{\citenamefont {Bender}\ \emph {et~al.}(2010)\citenamefont {Bender},
  \citenamefont {Courteille}, \citenamefont {Marzok}, \citenamefont
  {Zimmermann},\ and\ \citenamefont {Slama}}]{Bender2010}%
  \BibitemOpen
  \bibfield  {author} {\bibinfo {author} {\bibfnamefont {H.}~\bibnamefont
  {Bender}}, \bibinfo {author} {\bibfnamefont {P.~W.}\ \bibnamefont
  {Courteille}}, \bibinfo {author} {\bibfnamefont {C.}~\bibnamefont {Marzok}},
  \bibinfo {author} {\bibfnamefont {C.}~\bibnamefont {Zimmermann}}, \ and\
  \bibinfo {author} {\bibfnamefont {S.}~\bibnamefont {Slama}},\ }\href
  {\doibase 10.1103/PhysRevLett.104.083201} {\bibfield  {journal} {\bibinfo
  {journal} {Physical Review Letters}\ }\textbf {\bibinfo {volume} {104}},\
  \bibinfo {pages} {083201} (\bibinfo {year} {2010})}\BibitemShut {NoStop}%
\bibitem [{\citenamefont {Chang}\ \emph {et~al.}(2014)\citenamefont {Chang},
  \citenamefont {Sinha}, \citenamefont {Taylor},\ and\ \citenamefont
  {Kimble}}]{Chang2014}%
  \BibitemOpen
  \bibfield  {author} {\bibinfo {author} {\bibfnamefont {D.~E.}\ \bibnamefont
  {Chang}}, \bibinfo {author} {\bibfnamefont {K.}~\bibnamefont {Sinha}},
  \bibinfo {author} {\bibfnamefont {J.~M.}\ \bibnamefont {Taylor}}, \ and\
  \bibinfo {author} {\bibfnamefont {H.~J.}\ \bibnamefont {Kimble}},\ }\href
  {\doibase 10.1038/ncomms5343} {\bibfield  {journal} {\bibinfo  {journal}
  {Nature Communications}\ }\textbf {\bibinfo {volume} {5}} (\bibinfo {year}
  {2014}),\ 10.1038/ncomms5343}\BibitemShut {NoStop}%
\bibitem [{\citenamefont {Fuchs}\ \emph {et~al.}(2017)\citenamefont {Fuchs},
  \citenamefont {Bennett}, \citenamefont {Krems},\ and\ \citenamefont
  {Buhmann}}]{Fuchs2017c}%
  \BibitemOpen
  \bibfield  {author} {\bibinfo {author} {\bibfnamefont {S.}~\bibnamefont
  {Fuchs}}, \bibinfo {author} {\bibfnamefont {R.}~\bibnamefont {Bennett}},
  \bibinfo {author} {\bibfnamefont {R.~V.}\ \bibnamefont {Krems}}, \ and\
  \bibinfo {author} {\bibfnamefont {S.~Y.}\ \bibnamefont {Buhmann}},\ }\href
  {http://arxiv.org/abs/1711.10383} {\  (\bibinfo {year} {2017})},\ \Eprint
  {http://arxiv.org/abs/1711.10383} {arXiv:1711.10383} \BibitemShut {NoStop}%
\bibitem [{\citenamefont {Carniglia}\ and\ \citenamefont
  {Mandel}(1971)}]{Carniglia1971}%
  \BibitemOpen
  \bibfield  {author} {\bibinfo {author} {\bibfnamefont {C.~K.}\ \bibnamefont
  {Carniglia}}\ and\ \bibinfo {author} {\bibfnamefont {L.}~\bibnamefont
  {Mandel}},\ }\href {\doibase 10.1103/PhysRevD.3.280} {\bibfield  {journal}
  {\bibinfo  {journal} {Physical Review D}\ }\textbf {\bibinfo {volume} {3}},\
  \bibinfo {pages} {280} (\bibinfo {year} {1971})}\BibitemShut {NoStop}%
\bibitem [{\citenamefont {Derjaguin}(1934)}]{Derjaguin1934}%
  \BibitemOpen
  \bibfield  {author} {\bibinfo {author} {\bibfnamefont {B.}~\bibnamefont
  {Derjaguin}},\ }\href {\doibase 10.1007/BF01433225} {\bibfield  {journal}
  {\bibinfo  {journal} {Kolloid-Zeitschrift}\ }\textbf {\bibinfo {volume}
  {69}},\ \bibinfo {pages} {155} (\bibinfo {year} {1934})}\BibitemShut
  {NoStop}%
\bibitem [{\citenamefont {Lussange}\ \emph {et~al.}(2012)\citenamefont
  {Lussange}, \citenamefont {Gu{\'{e}}rout},\ and\ \citenamefont
  {Lambrecht}}]{Lussange2012}%
  \BibitemOpen
  \bibfield  {author} {\bibinfo {author} {\bibfnamefont {J.}~\bibnamefont
  {Lussange}}, \bibinfo {author} {\bibfnamefont {R.}~\bibnamefont
  {Gu{\'{e}}rout}}, \ and\ \bibinfo {author} {\bibfnamefont {A.}~\bibnamefont
  {Lambrecht}},\ }\href {\doibase 10.1103/PhysRevA.86.062502} {\bibfield
  {journal} {\bibinfo  {journal} {Physical Review A}\ }\textbf {\bibinfo
  {volume} {86}},\ \bibinfo {pages} {062502} (\bibinfo {year}
  {2012})}\BibitemShut {NoStop}%
\bibitem [{\citenamefont {Eberlein}\ and\ \citenamefont
  {Robaschik}(2006)}]{Eberlein2006}%
  \BibitemOpen
  \bibfield  {author} {\bibinfo {author} {\bibfnamefont {C.}~\bibnamefont
  {Eberlein}}\ and\ \bibinfo {author} {\bibfnamefont {D.}~\bibnamefont
  {Robaschik}},\ }\href {\doibase 10.1103/PhysRevD.73.025009} {\bibfield
  {journal} {\bibinfo  {journal} {Physical Review D}\ }\textbf {\bibinfo
  {volume} {73}},\ \bibinfo {pages} {025009} (\bibinfo {year}
  {2006})}\BibitemShut {NoStop}%
\bibitem [{\citenamefont {Bennett}\ and\ \citenamefont
  {Eberlein}(2012)}]{Bennett2012}%
  \BibitemOpen
  \bibfield  {author} {\bibinfo {author} {\bibfnamefont {R.}~\bibnamefont
  {Bennett}}\ and\ \bibinfo {author} {\bibfnamefont {C.}~\bibnamefont
  {Eberlein}},\ }\href {\doibase 10.1103/PhysRevA.86.062505} {\bibfield
  {journal} {\bibinfo  {journal} {Physical Review A}\ }\textbf {\bibinfo
  {volume} {86}},\ \bibinfo {pages} {062505} (\bibinfo {year}
  {2012})}\BibitemShut {NoStop}%
\bibitem [{\citenamefont {Kramida}\ \emph {et~al.}(2017)\citenamefont
  {Kramida}, \citenamefont {Ralchenko},\ and\ \citenamefont
  {Reader}}]{NIST_ASD}%
  \BibitemOpen
  \bibfield  {author} {\bibinfo {author} {\bibfnamefont {A.}~\bibnamefont
  {Kramida}}, \bibinfo {author} {\bibfnamefont {Y.}~\bibnamefont {Ralchenko}},
  \ and\ \bibinfo {author} {\bibfnamefont {J.}~\bibnamefont {Reader}},\ }\href
  {https://physics.nist.gov/asd} {\enquote {\bibinfo {title} {{NIST Atomic
  Spectra Database (ver. 5.5.1), [Online]. Available:
  https://physics.nist.gov/asd}},}\ }\bibinfo {howpublished} {National
  Institute of Standards and Technology, Gaithersburg, MD.} (\bibinfo {year}
  {2017})\BibitemShut {NoStop}%
\end{thebibliography}
\end{document}